\documentstyle[12pt]{article}
\newcommand{\ltwid}{\raise.3ex\hbox{$<$\kern-.75em\lower1ex\hbox{$\sim
$}}}

\def\Journal#1#2#3#4{{#1} {\bf #2}, #3 (#4)}

\def\NPB{{\em Nucl. Phys.} B}
\def\PLB{{\em Phys. Lett.}  B}
\def\PRL{\em Phys. Rev. Lett.}
\def\PRD{{\em Phys. Rev.} D}
\def\ZPC{{\em Z. Phys.} C}
\def\ZPA{{\em Z. Phys.} A}

\def\be{\begin{equation}}
\def\ee{\end{equation}}
\def\bea{\begin{eqnarray}}
\def\eea{\end{eqnarray}}

\begin{document}
\begin{center}
{\large \bf Magnetic Catalysis of Chiral Symmetry Breaking in QED at Finite
Temperature\footnote{\it To be published in Proceedings of
APCTP-ICTP Joint International conference on "Recent Developments in
Non-perturbative Methods", held on May 26-30, 1997 at Seoul, Korea.}}\\[6mm]

V.~P.~Gusynin\footnote{E-mail: vgusynin@gluk.apc.org}\\
{\it Bogolyubov Institute for Theoretical Physics,
252143 Kiev, Ukraine}\\
September 14, 1997\\[8mm]
\end{center}

\begin{abstract}
The catalysis of chiral symmetry breaking by a magnetic field in the massless
weak-coupling phase of QED is studied. The dynamical mass of a fermion (energy
gap in the fermion spectrum) is shown to depend essentially nonanalytically on
the renormalized coupling constant $\alpha$ in a strong magnetic field.
The temperature of the symmetry restoration is calculated
analytically as $T_c\approx m_{dyn}$, where $m_{dyn}$ is the dynamical mass of
a fermion at zero temperature.
\end{abstract}
\newpage

In this talk I will discuss dynamical chiral symmetry breaking in a
magnetic field and at finite temperature. The talk is based on the recent
papers with V.Miransky and I.Shovkovy \cite{GMS,NP,GS}.

The dynamics of fermions in an external constant magnetic field in QED
was considered by Schwinger long ago \cite{Sch}. In that work,
while the interaction with the external field was considered in all orders
in the coupling constant, the quantum dynamics was treated perturbatively.
There is no dynamical chiral symmetry breaking in QED in this approximation.
Also, chiral symmetry breaking is not manifested in the weak coupling phase
of QED in the absense of a magnetic field, even if it is treated nonperturbatively.
We will show that a constant magnetic field $B$ changes the situation drastically,
namely, it leads to dynamical chiral symmetry breaking in QED for any arbitrary
weak interaction. The essence of this magnetic catalysis is that electrons
are effectively 1+1 dimensional when their energy is much less than the
Landau gap $\sqrt{|eB|}$ what was pointed out recently in Refs.\cite{GMS1,GMS2}.
The lowest Landau level (LLL) plays here the role
similar to that of the Fermi surface in the BCS theory of superconductivity,
leading to dimensional reduction in dynamics of fermion pairing.

The dynamical mass of fermions (energy gap in
the fermion spectrum) is:
\begin{equation}
m_{dyn}\simeq C\sqrt{eB}\exp{\left[-\left(\frac{\pi}{\alpha}\right)^{1/2}
\right]},
\label{mdyn}
\end{equation}
where the constant $C$ is of order one and $\alpha=e^2 /4\pi$ is the
renormalized coupling constant.

The effect of magnetic catalysis was studied also in Nambu-Jona-Lasino (NJL)
models in 2+1 \cite{GMS1,Klim} and 3+1 dimensions \cite{GMS2}, it was
extended to the case of external non-abelian chromomagnetic fields and
finite temperatures \cite{Igor,Ebert}, curved spacetime \cite{Odintsov},
as well as to the supersymmetric NJL
model \cite{Elias}, confirming the universality of the phenomenon.

We emphasize that we will work in the conventional, weak coupling, phase of
QED. That is, the bare coupling $\alpha^{(0)}$, relating to the scale
$\mu=\Lambda$, where $\Lambda$ is an ultraviolet cutoff, is assumed to be
small, $\alpha^{(0)}\ll 1$. Then, because of infrared freedom in QED,
interactions in the theory are weak at all scales and, as a result, the
treatment of the nonperturbative dynamics is reliable.

An important question of the chiral symmetry restoration in QED at finite
temperature was addressed in the recent work of Lee, Leung and Ng \cite{Leung}
who have obtained for the critical temperature $T_c\simeq\frac{\alpha}{\pi^2}
\sqrt{2\pi|eB|}$. However, their $T_c$ can be considered only as a rough upper
estimate. We will show that the correct estimate for the critical temperature
is $T_c\approx m_{dyn}$ with $m_{dyn}$ given by (\ref{mdyn}).

The Lagrangian density of massless QED in a magnetic field is:
\begin{equation}
{\cal L} = -{1\over4}F^{\mu\nu}F_{\mu\nu} + \frac{1}{2} \left[\bar{\psi},
(i\gamma^\mu D_\mu)\psi\right],
\label{eq:lag}
\end{equation}
where the covariant derivative $D_\mu$ is
\begin{equation}
D_\mu=\partial_\mu-ie(A^{ext}_\mu+A_\mu),\qquad A^{ext}_\mu =
\left(0,-\frac{B}{2} x_2 ,\frac{B}{2} x_1,0\right).
\label{eq:vecA}
\end{equation}

The Lagrangian density (\ref{eq:lag}) is chiral invariant
( we will not discuss the dynamics related to the anomalous
current $j_{5\mu}$, in any case it is not manifested in quenched approximation
dealt with in present consideration). When the chiral symmetry is broken there
appears a gapless NG boson composed of fermion and antifermion. The dynamical
mass for a fermion can be defined by considering the Bethe-Salpeter (BS)
equation for NG boson \cite{GMS,NP} or the Schwinger-Dyson (SD) equation for
the dynamical mass function  \cite{Ng,Hong}.
We consider here the homogeneous BS equation for  NG bound state which takes
the form:
\begin{eqnarray}
\chi_{\alpha\beta}(x,y;P)&=&-i\int d^4x_1 d^4y_1 d^4x_2 d^4y_2 G_{\alpha
\alpha_{1}}(x,x_1)K_{\alpha_{1}\beta_{1};\alpha_{2}\beta_{2}}(x_1y_1,x_2y_2)
\nonumber\\
&\cdot&\chi_{\alpha_{2}\beta_{2}}(x_2,y_2;P)G_{\beta_1\beta}(y_1,y),
\end{eqnarray}
where the BS wave function $\chi_{\alpha\beta}(x,y;P)=\langle 0|T\psi_\alpha(x)
{\bar\psi}_\beta(y)|P \rangle$, and the fermion propagator $G_{\alpha\beta}
(x,y)=\langle 0|T\psi_\alpha(x){\bar\psi}_\beta(y)|0 \rangle$.
Note that though the external field $A_\mu^{ext}$ (\ref{eq:vecA}) breaks the
conventional translation invariance, the total momentum $P$ is a good,
conserved, quantum number for neutral channels \cite{Avron}, in particular,
for this NG boson. Since, as will be shown below, the NG boson is
formed in the infrared region, where the QED coupling  is weak, one
can use the BS kernel in leading order in $\alpha$ :
\begin{eqnarray}
K_{\alpha_{1}\beta_{1};\alpha_{2}\beta_{2}}(x_1y_1,x_2y_2)=-4\pi i\alpha
\gamma^\mu_{\alpha_1\alpha_2}\gamma^\nu_{\beta_2\beta_1}{\cal D}_{\mu\nu}
(y_2-x_2)\delta(x_1-x_2)\delta(y_1-y_2),
\label{eq:ker}
\end{eqnarray}
where the photon propagator
\begin{equation}
{\cal D}_{\mu\nu}(x)=\frac{-i}{(2\pi)^4}\int d^4k e^{ikx} \left(g_{\mu\nu}-
\lambda\frac{k_\mu k_\nu}{k^2}\right)\frac{1}{k^2}\label{eq:dmunu}
\end{equation}
($\lambda$ is a gauge parameter).
Then the BS equation takes the form:
\begin{eqnarray}
\chi_{\alpha\beta}(x,y;P)&=&-4\pi\alpha\int d^4x_1 d^4y_1 \bigg[S(x,x_1)
\gamma^\mu\chi(x_1,y_1;P)\gamma^\nu S(y_1,y)\bigg]_{\alpha\beta}
{\cal D}_{\mu\nu}(y_1-x_1),
\label{eq:bs1}
\end{eqnarray}
where, since we are working to the lowest order in $\alpha$, the full
fermion propagator $G(x,y)$ is replaced by the propagator $S$ of a free
fermion (with the mass $m=m_{dyn}$) in a magnetic field \cite{Sch}:
\begin{equation}
S(x,y) = \exp \left(\frac{ie}{2}(x-y)^\mu A_\mu^{ext}
(x+y)\right)\tilde{S}(x-y),
\end{equation}
where the Fourier transform of $\tilde{S}$ is
\begin{eqnarray}
& &\tilde{S}(k) =\int\limits^\infty_0 ds \exp \left[is\left(k^2_0-k^2_3
-k^2_{\perp}\frac{\tan(eBs)}{eBs}-m_{dyn}^2\right)\right] \nonumber \\
& &\cdot\left[(k^0\gamma^0-k^3\gamma^3+m_{dyn})(1+\gamma^1\gamma^2\tan(eBs))
-k_{\perp}\gamma_{\perp}(1+\tan^2(eBs))\right].
\end{eqnarray}
Here $k_{\perp}=(k_1,k_2)$, $\gamma_{\perp}=
(\gamma_1,\gamma_2)$. Using the variables, the center-of-mass coordinate,
$R=(x+y)/2$, and the relative coordinate, $r=x-y$, equation (\ref{eq:bs1})
can be rewritten as
\begin{eqnarray}
& &\tilde{\chi}_{\alpha\beta}(R,r;P)=-4\pi\alpha\int d^4R_1 d^4r_1 \left[
\tilde{S}\left(R-R_1+\frac{r-r_1}{2}\right)\gamma^\mu\tilde{\chi}(R_1,r_1;P)
\gamma^\nu\right.\nonumber\\
& &\cdot\left.\tilde{S}\left(\frac{r-r_1}{2}-R+R_1\right)
\right]_{\alpha\beta}\!{\cal D}_{\mu\nu}(-r_1)\exp\left[-ie(r+r_1)^\mu A_
\mu^{ext}(R-R_1)\right]\!.
\label{eq:bs12}
\end{eqnarray}
Here the function $\tilde{\chi}_{\alpha\beta}(R,r;P)$ is defined from the
equation
\begin{equation}
\chi_{\alpha\beta}(x,y;P)=\langle 0|T\psi_\alpha(x){\bar\psi}_\beta(y)|P
\rangle= \exp[ier^\mu A_\mu^{ext}(R)]\tilde{\chi}_{\alpha\beta}(R,r;P).
\label{eq:bsfun}
\end{equation}
It is important that the effect of translation symmetry breaking by the
external field is factorized in the phase factor in Eq.(\ref{eq:bsfun})
and equation (\ref{eq:bs12}) admits a translation invariant solution,
$\tilde{\chi}(R,r;P)=\exp(-iPR)\tilde{\chi}(r;P)$. Henceforth we will
consider the case $P=0$, corresponding to NG bound state.
Then, transforming this equation into momentum space, we get:
\begin{eqnarray}
& &\tilde{\chi}_{\alpha\beta}(p)=-4\pi\alpha\int \frac{d^2q_{\perp} d^2R_
{\perp}d^2k_{\perp} d^2k_{\parallel}}{(2\pi)^6} \exp(-iq_{\perp} R_{\perp})
\left[\tilde{S}\left(p_{\parallel}, p_{\perp}+e A^{ext}(R_{\perp})
+\frac{q_{\perp}}{2}\right)\right.\nonumber\\
& &\cdot\left.\gamma^\mu\tilde{\chi}(k)\gamma^\nu
\tilde{S}(p_{\parallel},p_{\perp}+e A^{ext}(R_{\perp})-\frac{q_{\perp}}{2})
\right]_{\alpha\beta}{\cal D}_{\mu\nu}(k_{\parallel}-p_{\parallel},
k_{\perp}-p_{\perp}-2e A^{ext}(R_{\perp})),
\label{eq:bs2}
\end{eqnarray}
where $p_{\parallel}\equiv (p^0,p^3)$, $p_{\perp}\equiv (p^1,p^2)$.

The crucial point for the further analysis will be the assumption that
$m_{dyn}\ll \sqrt{|eB|}$ and that the region mostly responsible for
generating the mass is the infrared region with $k\ll\sqrt{|eB|}$.
The assumption allows us to replace the propagator
$\tilde{S}$ in Eq.(\ref{eq:bs2}) by that projected into LLL.
In order to show this, we recall that the energy spectrum of
fermions with $m=m_{dyn}$ in a magnetic field is
\begin{equation}
E_n(p_3) = \pm\sqrt{m^2_{dyn}+2|eB|n+p^2_3},\ \qquad  n=0,1,2,\dots
\end{equation}
(the Landau levels). The  propagator $\tilde{S}(p)$ can be decomposed
over the Landau level poles as follows \cite{GMS2,Cho} :
\begin{equation}
\tilde{S}(p)=i \exp\left(-\frac{p^2_{\perp}}{|eB|}\right)
\sum_{n=0}^{\infty}(-1)^n\frac{D_n(eB,p)}{p_0^2-p_3^2-m_{dyn}^2-2|eB|n}
\label{eq:poles}
\end{equation}
with
\begin{eqnarray*}
D_n(eB,p)&=&(p^0\gamma^0-p^3\gamma^3+m_{dyn})\left[(1-i\gamma^1\gamma^2
{\rm sign}(eB))L_n\left(2\frac{p^2_{\perp}}{|eB|}\right)\right.\\
&-&\left.(1+i\gamma^1\gamma^2{\rm sign}(eB))
L_{n-1}\left(2\frac{p^2_{\perp}}{|eB|}\right)\right]
+4(p^1\gamma^1+p^2\gamma^2)L_{n-1}^1\left(2\frac{p^2_{\perp}}{|eB|}\right),
\end{eqnarray*}
where $L_n(x)$ are the generalized Laguerre polynomials ($L_n\equiv L_n^0$,
$L_{-1}^{\alpha}(x)=0$).  Eq.(\ref{eq:poles})
implies that at $p^2_{\parallel}, m_{dyn}\ll \sqrt{|eB|}$, the LLL with $n=0$
dominates and we can take
\begin{eqnarray}
\tilde{S}(p)&\simeq &i\exp(-\frac{p^2_{\perp}}{|eB|})\frac{\hat{p}_{\parallel}
+m_{dyn}}{p^2_{\parallel}-m^2_{dyn}}(1-i\gamma^1\gamma^2{\rm sign}(eB)),
\label{eq:0lev}
\end{eqnarray}
where $\hat{p}_{\parallel}=p^0\gamma^0-p^3\gamma^3$ and
$\hat{p}_{\parallel}^2=(p^0)^2-(p^3)^2$, and Eq.(\ref{eq:bs2})
transforms into the following one (further for concretness we assume $eB>0$):
\begin{eqnarray}
\rho(p_{\parallel},p_{\perp})&=&\frac{2\alpha l^2}{(2\pi)^4}
e^{-l^2 p^2_{\perp}}\int d^2A_{\perp} d^2k_{\perp}d^2k_{\parallel}
e^{-l^2A^2_{\perp}}(1-i\gamma^1\gamma^2)\gamma^\mu
\frac{\hat{k}_{\parallel}+m_{dyn}}{k^2_{\parallel}-m^2_{dyn}} \nonumber\\
&\cdot&\rho(k_{\parallel}, k_{\perp})\frac{\hat{k}_{\parallel}+m_{dyn}}
{k^2_{\parallel}-m^2_{dyn}}\gamma^\nu (1-i\gamma^1\gamma^2)
{\cal D}_{\mu\nu}(k_{\parallel}-p_{\parallel}, k_{\perp}- A_{\perp}),
\label{eq:rho}
\end{eqnarray}
where
\begin{equation}
\rho(p_{\parallel}, p_{\perp})=(\hat{p}_{\parallel}-m_{dyn})
\tilde{\chi}(p)(\hat{p}_{\parallel}-m_{dyn}),
\end{equation}
and $l=|eB|^{-1/2}$ is the magnetic length.
Eq.(\ref{eq:rho}) implies that $\rho(p_{\parallel}, p_{\perp})=
\exp(-l^2 p^2_{\perp})
\varphi(p_{\parallel})$, where $\varphi(p_{\parallel})$ satisfies
the equation:
\begin{eqnarray}
\varphi(p_{\parallel})&=&\frac{\pi\alpha}{(2\pi)^4}\int d^2k_{\parallel}
(1-i\gamma^1\gamma^2)\gamma^\mu\frac{\hat{k}_{\parallel}+m_{dyn}}
{k^2_{\parallel}-m^2_{dyn}}\varphi(k_{\parallel}) \frac{\hat{k}_{
\parallel}+m_{dyn}}{k^2_{\parallel}-m^2_{dyn}}\gamma^\nu (1-i\gamma^1
\gamma^2)\nonumber\\
&\cdot&{\cal D}^{\parallel}_{\mu\nu}(k_{\parallel}-p_{\parallel}),
\label{eq:phi}
\end{eqnarray}
\begin{equation}
{\cal D}^{\parallel}_{\mu\nu}(k_{\parallel}-p_{\parallel})=
\int d^2k_{\perp}\exp(-\frac{l^2 k^2_{\perp}}{2})
{\cal D}_{\mu\nu}(k_{\parallel}-p_{\parallel}, k_{\perp}).
\end{equation}
Thus the BS equation has been reduced to a two--dimensional integral
equation. Of course, this fact reflects the two--dimensional character of
the dynamics of electrons from LLL, that can be explicitly
read from Eq.(\ref{eq:0lev}).

We emphasize that the dimensional reduction  in a magnetic field does not
affect the dynamics of the center of mass of {\em neutral} bound
states (in particular, this NG boson). Indeed, the reduction
$3+1\to 1+1$ in the fermion propagator, in the infrared region, reflects
the fact that the motion of {\em charged} particles is restricted in
directions perpendicular to the magnetic field. Since there is no such
a restriction for the motion of the center of mass of neutral particles,
their propagator must have a $(3+1)$--dimensional form. This
fact was explicitly shown for neutral bound states in NJL model in a
magnetic field, in $1/N_c$ expansion \cite{GMS2}, and for neutral excitations
in nonrelativistic systems \cite{Cond}.  This in particular implies that,
notwithstanding the dimensional reduction in a magnetic field, the phenomenon
of spontaneous chiral symmetry breaking in QED does not contradict to the
Mermin--Wagner--Coleman theorem \cite{Mermin} forbidding the spontaneous
breakdown of continuous symmetries at $D=1+1$.

Henceforth we will use Euclidean space with $k_4=-ik^0$. In order to define
the matrix structure of the wave function $\varphi(p_{\parallel})$ of the
NG boson, note that, in a magnetic field, there is the symmetry $SO(2)
\times SO(2)\times {\cal P}$, where the $SO(2)\times SO(2)$ is connected
with rotations in $x_1$--$x_2$ and $x_3$--$x_4$ planes and ${\cal P}$ is
the inversion transformation
$x_3\to -x_3$ (under which a fermion field transforms as $\psi\to i\gamma_5
\gamma_3\psi$). This symmetry implies that the function
$\varphi(p_{\parallel})$ takes the form:
\begin{equation}
\varphi(p_{\parallel})=\gamma_5 (A+i\gamma_1\gamma_2B+\hat{p}_{\parallel}C+
i\gamma_1\gamma_2\hat{p}_{\parallel}D),   \label{eq:phii}
\end{equation}
where $\hat{p}_{\parallel}=p_3\gamma_3+p_4\gamma_4$ (with $\gamma_\mu$ anti
-Hermitian in Euclidean space) and $A,B,C$ and $D$ are functions of $p^2_
{\parallel}$.

In Feynman gauge
\begin{eqnarray*}
{\cal D}^{\parallel}_{\mu\nu}(k_{\parallel}-p_{\parallel})=\delta_{\mu\nu}\pi
\int\limits^{\infty}_{0}\frac{dx \exp(-l^2x/2)}{(k_{\parallel}-
p_{\parallel})^2+x},
\end{eqnarray*}
and, substituting expansion (\ref{eq:phii}) into equation (\ref{eq:phi}),
we find that $B=-A$, $C=D=0$, {\em i.e.},
$\varphi(p_{\parallel})=A\gamma_5(1-i\gamma_1\gamma_2)$,
and the function $A$ satisfies the equation
\begin{equation}
A(p)=\frac{\alpha}{2\pi^2}\int\frac{d^2k A(k)}{k^2+m^2_{dyn}}
\int\limits^{\infty}_{0} \frac{dx \exp(-xl^2/2)}{(k-p)^2+x}
\label{eq:Ap}
\end{equation}
(henceforth we will omit the symbol $\parallel$ from $p$ and $k$).
Eq.(\ref{eq:Ap}) coincides with the equation for the dynamical fermion mass
function obtained with the help of SD equation for a fermion propagator in
\cite{Leung,Hong}.
As was shown in \cite{NP} (see
Appendix C), in the case of weak coupling $\alpha$, the function $A(p)$ remains almost
constant in the range of momenta $0<p^2\ltwid 1/l^2$ and decays like $1/p^2$
outside that region. To get an estimate for $m_{dyn}$ at $\alpha<<1$, we set
the external momentum to be zero and notice that the main contribution in the
integral on the right hand side of Eq.(\ref{eq:Ap})
is formed in the infrared region with $k^2\ltwid 1/l^2$. The latter validates
in its turn the substitution $A(k)\rightarrow A(0)$ in the integrand of (\ref
{eq:Ap}), and we finally come to the following gap equation:
\begin{equation}
A(0)\simeq \frac{\alpha}{2\pi^2}A(0)\int\frac{d^2k}{k^2+m^2_{dyn}}
\int\limits^{\infty}_{0} \frac{dx \exp(-l^2x/2)}{k^2+x},
\label{gapeq}
\end{equation}
{\em i.e.},
\begin{equation}
1\simeq\frac{\alpha}{2\pi}\int\limits_0^\infty\frac{dx\exp(-ax)}{x-1}\log x,
\qquad a\equiv\frac{m^2_{dyn}l^2}{2}.
\label{eqmdyn}
\end{equation}
The main contribution in (\ref{eqmdyn}) comes from the region $x\ltwid1/a$,
thus at $a<<1$ we get
\begin{equation}
1\simeq\frac{\alpha}{4\pi}\log^2\left(\frac{m^2_{dyn}l^2}{2}\right),
\end{equation}
therefrom the expression (\ref{mdyn}) for $m_{dyn}$ follows. The exponential
factor in $m_{dyn}$ displays the nonperturbative nature of this result. It
can be shown also that the expression (\ref{mdyn}) for the dynamical mass is
gauge invariant \cite{GMS}.

More accurate analysis \cite{NP}, which takes into account the momentum
dependence of the mass function, leads to the result
\begin{equation}
m_{dyn}\simeq C\sqrt{|eB|}\exp\left[-{\pi\over2}\sqrt{\frac{\pi}{2\alpha}}
\right].
\end{equation}
Notice that the ratio of the powers of this exponent and that in Eq.(\ref
{mdyn}) is $\pi/2\sqrt2\simeq1.1$, thus the approximation used above is
rather reliable.

To study chiral symmetry breaking in an external field at nonzero temperature
we use the imaginary time formalism. Now the analogue of the equation (\ref
{eq:Ap}) (with the replacement $m_{dyn}\rightarrow m^2(T)$ in the denominator)
reads
\begin{equation}
A(\omega_{n'},p)=\frac{\alpha}{\pi}T\sum_{n=-\infty}^\infty\int\limits_
{-\infty}^\infty\frac{dkA(\omega_n,k)}{\omega^2_n+k^2+m^2(T)}\int\limits_0^
\infty\frac{dx\exp(-l^2x/2)}{(\omega_n-\omega_{n'})^2+(k-p)^2+x},
\label{tempgap}
\end{equation}
where $\omega_n=\pi T(2n+1)$ are Matsubara frequencies.

If we now take $n'=0, p=0$ in the left hand side of Eq.(\ref{tempgap}) and
put $A(\omega_n,k)\approx A(\omega_0,0)=const$ in the integrand, we come to
the equation
\begin{equation}
1=\frac{\alpha}{\pi}T\sum_{n=-\infty}^\infty\int\limits_{-\infty}^\infty
\frac{dk}{\omega^2_n+k^2+m^2(T)}\int\limits_0^\infty\frac{dx\exp(-l^2x/2)}
{(\omega_n-\omega_0)^2+k^2+x}.
\label{tgap}
\end{equation}
After evaluating the sum in (\ref{tgap}), we are left with the equation
\bea
&&\hspace{-3mm}1=\frac{\alpha}{\pi} \int\limits_{0}^{\infty}
\int\limits^\infty_0
\frac{dk dx \exp[-\ell^2x/2]}{[(\pi T)^2 +x -m^2(T)]^2
+(2\pi T)^2 (k^2+m^2(T))}\left\{
\frac{(\pi T)^2 +x -m^2(T)}{\sqrt{k^2 +m^2(T)}}\right.\nonumber\\
&&\left.\cdot\tanh \left(\frac{\sqrt{k^2+m^2(T)}}{2T}\right)
+\frac{(\pi T)^2 +m^2(T) -x}{\sqrt{k^2+x}}
\coth \left(\frac{\sqrt{k^2+x}}{2T}\right)
\right\}.
\label{eq:our3}
\eea
The equation for the critical temperature is obtained from
(\ref{eq:our3}) putting $m(T_c)=0$:
\be
1=\frac{\alpha}{\pi}
\int\limits_{0}^{\infty} \int\limits^\infty_0
\frac{dk dx e^{-2x(\pi T_c\ell)^2}}{[1/4 +x]^2 +k^2}
\left\{ \frac{1/4 +x}{k} \tanh \left(\pi k\right)
+\frac{1/4 -x}{\sqrt{k^2+x}} \coth \left(\pi\sqrt{k^2+x}\right)
\right\},
\label{eq:t_c}
\ee
where we also switched to dimensionless variables $x\to (2\pi
T_{c})^2 x$ and $k\to 2\pi T_{c} k$.

By assuming smallness of the critical temperature in comparison
with the scale put by the magnetic field, $T_c\ell\ll 1$, we see
that the double logarithmic in field contribution in
Eq.(\ref{eq:t_c}) comes from the region $0< x \ltwid 1/2(\pi
T_c\ell)^2 $, $1/\pi \ltwid k<\infty$.  Simple estimate gives:
\begin{eqnarray}
& &1\simeq\frac{\alpha}{\pi} \int\limits_0^{1/2(\pi T_{c}\ell)^2}
dx \int\limits_{1/\pi}^{\infty} \frac{dk}{[1/4 +x]^2 +k^2}\left[
\frac{1/4 +x}{k}+\frac{1/4-x}{\sqrt{k^2+x}}\right]
\nonumber\\
& &\simeq\frac{\alpha}{\pi} \int\limits_0^{1/2(\pi T_c\ell)^2}dx
\left[\frac{1}{2(1/4 +x)} \log\left(1+(1/4 +x)^2\pi^2\right)
     +\frac{1/4-x}{(1/4+x)|1/4-x|}\right.\nonumber\\
& &\cdot\left.\log\frac{(1/4+x+|1/4-x|)\sqrt{1/\pi^2
+(1/4+x)^2}}
{(1/4+x)\sqrt{1/\pi^2+x}+|1/4-x|/\pi}\right]\simeq
\frac{\alpha}{4\pi}\log^2\left[\frac{1}{2(\pi T_c\ell)^2}\right].
\label{integ}
\end{eqnarray}
Thus, for the critical temperature, we obtain the estimate:
\begin{eqnarray}
T_{c}\approx \sqrt{|eB|}\exp\left[-\sqrt{\frac{\pi}{\alpha}}\right]
\approx m_{dyn}(T=0),
\label{t_c-m_dyn}
\end{eqnarray}
where $m_{dyn}$ is given by (\ref{mdyn}). The relationship
$T_c\approx m_{dyn}$ between the critical temperature and the zero
temperature fermion mass was obtained also in NJL model in (2+1)-
and (3+1)-dimensions \cite{GMS2,Ebert}. The constant $C$, in the
relation $T_c=Cm_{dyn}$, is of order one and can be calculated
numerically.  We note the photon thermal mass, which is of the order
of $\sqrt\alpha T$ \cite{Weldon}, cannot change our result for the
critical temperature. As is easy to check, the only effect of
taking it into account will be the shift in $x$ for a constant of
the order of $\alpha$ in the integrand of (\ref{integ}). However,
such a shift is absolutely irrelevant for our estimate (\ref
{t_c-m_dyn}).
On the other hand, there are relevant one-loop contributions
in the photon propagator due to a magnetic field. Taking them into
account (that corresponds to the so-called improved ladder
approximation), we get an expression for $m_{dyn}$ of the form
(\ref{mdyn}) with the replacement $\alpha\rightarrow\alpha/2$ \cite
{NP}.

In conclusion, let us discuss possible applications of this effect.
One potential application is the interpretation of the results of
the GSI heavy--ion scattering experiments \cite{Sal} in which narrow
peaks are seen in the energy spectra of emitted $e^{+}e^{-}$ pairs.
One proposed explanation \cite{Celenza} is that a very strong
electromagnetic field, created by the heavy ions, induces a phase
transition in QED to a phase with spontaneous chiral symmetry breaking.
The observed peaks are due to the decay of positronium--like states in
this phase. The catalysis of chiral symmetry breaking by a magnetic
field in QED, discussed in this report, can serve as a toy example of
such a phenomenon. In order to get a more realistic model, it would
be interesting to extend this analysis to non--constant background fields

An interesting application of the magnetic catalysis in astrophysics,
such as the cooling process of neutron stars due to enhanced Primakoff
process in high magnetic field, was mentioned recently in \cite{Hongcomm}.

Yet another application of the effect is connected with the role of
chromomagnetic backgrounds imitating gluon condensate in the QCD vacuum
\cite{Igor,Ebert,Smilga}.

\vspace*{-2pt}
\section*{Acknowledgments}
I am grateful to the members of the Department of Physics of the
Pusan National University, especially  Dr.D.K.Hong, for their hospitality
during my stay there. The visit was supported by Basic Science Research
Program, Ministry of Education of Korea, 1996 (BSRI-96-2413). I would
like to thank the organizers of APCTP-ICTP Joint International conference
on "Recent Developments in Non-perturbative Methods", held on May 26-30,
1997 at Seoul, Korea, for invitation to take part in the conference and
support.  This study is supported by Foundation of Fundamental Researches
of Ministry of Sciences of the Ukraine under grant N 2.5.1/003 and Swiss
grant CEEC/NIS/96-98/7 IP 051219.


\end{document}